\documentclass[aps,prb,reprint,showpacs]{revtex4-1}
\usepackage[dvipdfmx]{graphicx}
\usepackage{amsmath,amssymb}
\usepackage{color}
\usepackage{bm}
\usepackage{hyperref}
\usepackage{ulem}

\begin{document}

\title{Microscopic theory of spin transport at the interface \\
between a superconductor and a ferromagnetic insulator}

\author{T. Kato$^{1}$, Y. Ohnuma$^{2}$, M. Matsuo$^{2}$, J. Rech$^{3}$, T. Jonckheere$^{3}$, T. Martin$^{3}$}
\affiliation{%
${^1}$Institute for Solid State Physics, The University of Tokyo, Kashiwa, Japan\\
${^2}$Kavli Institute for Theoretical Sciences, University of Chinese Academy of Sciences, Beijing, China \\
${^3}$Aix Marseille Univ, Universit\'e de Toulon, CNRS, CPT, Marseille, France 
}
\date{\today}

\begin{abstract}
We theoretically investigate spin transport at the interface between a ferromagnetic insulator (FI) and a superconductor (SC).
Considering a simple FI-SC interface model, we derive formulas for the spin current and spin-current noise induced by microwave irradiation (spin pumping) or the temperature gradient (the spin Seebeck effect).
We show how the superconducting coherence factor affects the temperature dependence of the spin current.
We also calculate the spin-current noise in thermal equilibrium and in non-equilibrium states induced by the spin pumping, and compare them quantitatively for an yttrium-iron-garnet-NbN interface.

\end{abstract}
\maketitle 

\section{Introduction}

Spin transport in hybrid systems composed of superconductors (SCs) and ferromagnetic metals has been investigated for a long time~\cite{Tedrow70,Meservey94,Zutic04,Linder15}.
In a superconductor, charge and spin imbalances may have different characteristic length scales due to spin-charge separation~\cite{Kivelson90,Zhao95,Bergeret18}.
The interplay between superconductivity and magnetism also offers the potential for novel spintronic devices, in which fast logic operation can be performed with minimum Joule heating~\cite{Sarma01}.
One of the key ingredients there is the injection of spin-polarized carriers into SCs~\cite{Bergeret18,Hubler12,Quay13,Wolf13,Wakamura14}.
For conventional $s$-wave superconductors, spin injection is suppressed by opening a superconducting gap in the electronic spectrum.
Thermally excited quasiparticles in SC, however, can carry a spin current over long distances, as spin excitations in SCs have long lifetimes~\cite{Yamashita02,Takahashi03,Morten04,Morten05,Silaev15,Aikebaier18}.

There are several techniques for spin injection into SCs. 
Recently, it has been realized by taking advantage of the spin Seebeck effect (SSE) induced by a temperature gradient,~\cite{Uchida08,Jaworski10,Uchida10,Xiao10,Adachi11,Adachi13,Ohnuma17} or by applying a spin pumping (SP) protocol using ferromagnetic resonance (FMR) under microwave irradiation~\cite{Tserkovnyak02,Konig03,Saitoh06,Kajiwara10,Ohnuma14}.
The latter technique has successfully been used in experiments to realize spin injection from ferromagnetic metals into a SC~\cite{Jeon18a,Jeon18b,Yao18}.
These recent advances indicate a new path for spin injection into a wide class of SC materials.
Remarkably, spin-current injection from a ferromagnetic insulator (FI) into a superconductor has also been performed recently~\cite{Umeda18}, as revealed by the inverse spin Hall effect (ISHE)~\cite{Takahashi02,Takahashi08,Wakamura15}.
This last study opens up possible applications for novel superconducting spintronic devices using FI.

In contrast to progress in experiments, the spin current at the FI-SC interface has been studied theoretically, to our knowledge, only by Inoue et al.~\cite{Inoue17}
They have formulated the spin pumping signal in terms of the local spin susceptibility of the SC, and have shown that the signal is peaked below the transition temperature due to the coherence factor in the BCS theory.
In order to calculate the local spin susceptibility of the SC, they have employed the Abrikosov-Gor'kov theory for dirty SCs taking spin diffusion into account~\cite{Abrikosov62,Gorkov64,Fulde68}.
The dynamic spin susceptibility thus obtained is, however, correct only for small wavenumbers, whereas the local spin susceptibility, which involves all wavenumbers, is dominated by the large wavenumber contribution~\cite{Shastry94} (for details, see Appendix~\ref{app:ImpurityEffect}).
Therefore, although their discovery of the coherence peak in spin transport is remarkable, their theory is expected to be insufficient for a quantitative description of the spin-current generation.

\begin{figure}[tb]
\begin{center}
\includegraphics[width=6cm,pagebox=cropbox,clip]{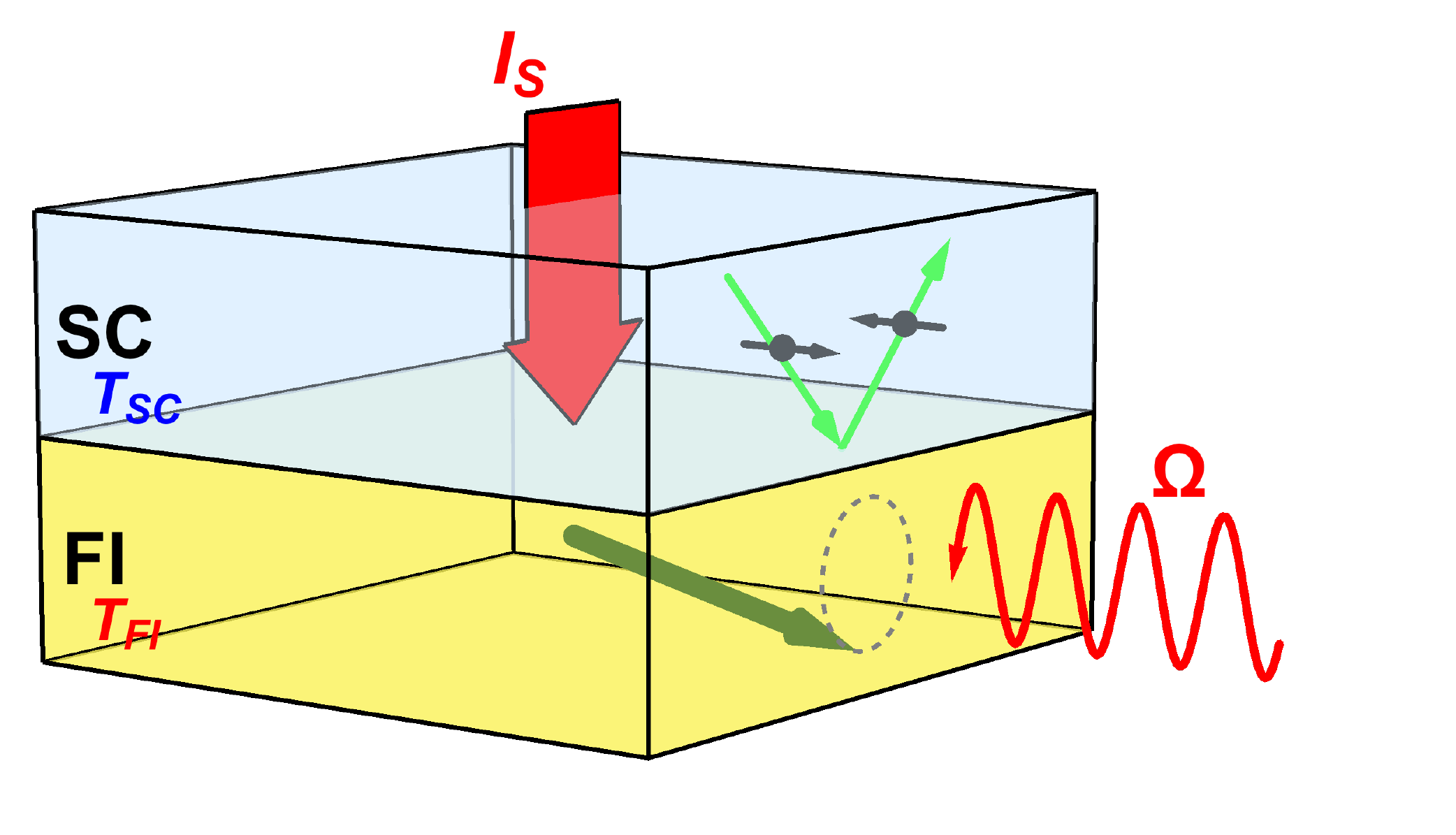}
\caption{Schematic picture of the FI-SC bilayer system.
A spin current $I_S$ is generated in the SC by spin pumping using an external microwave irradiation, or by spin Seebeck effect induced by a temperature gradient ($T_{\rm FI} \ne T_{\rm SC}$). The large green arrow in the FI illustrates the magnetization, which can precess due to the applied microwave at frequency $\Omega$. The arrows in the SC shows an example of electron reflection at the interface, with a spin flip due to the exchange interaction.}
\label{fig:setup}
\end{center}
\end{figure}

In this paper, we consider a bilayer system composed of a $s$-wave singlet SC and a FI as shown in Fig.~\ref{fig:setup}.
We formulate the spin current at the interface, and study its temperature dependence above and below the superconducting transition temperature.
We also discuss the noise power of the pure spin current following the theory developed by three of the present authors and one collaborator~\cite{Matsuo18}, and estimate it using the experimental parameters for the yttrium-iron-garnet(YIG)-NbN interface~\cite{Kajiwara10,Umeda18}.

This paper is organized as follows. 
We introduce the model for the FI-SC interface in Sec.~\ref{eq:model}, and derive dynamic spin susceptibilities in Sec.~\ref{sec:SpinSusceptibility}.
By using a second-order perturbative expansion with respect to the interface exchange coupling, we calculate the spin current and the spin-current noise in Sec.~\ref{sec:SpinCurrent} and Sec.~\ref{sec:SpinCurrentNoise}, respectively.
It should be stressed that we evaluate the spin current just at the FI-SC interface.
For experimental detection, one needs a nanostructure for converting the spin current into an electronic response, a mechanism which depends in general on details of spin relaxation in the superconductor.
We briefly discuss such a possible experimental setup for detecting the spin current in Sec.~\ref{sec:experiment}.
Finally, we summarize our results in Sec.~\ref{sec:summary}.
Detailed discussions on the impurity effect and the spin susceptibility of the SC are given in Appendix~\ref{app:ImpurityEffect} and \ref{app:BCS}, respectively.

\section{Model}
\label{eq:model}

The system Hamiltonian is given by $H = H_{\rm SC} + H_{\rm FI} + H_{\rm ex}$. 
The first term $H_{\rm SC}$ describes a bulk SC, and is given by the mean-field Hamiltonian
\begin{equation}
H_{\rm SC} = \sum_{\bm k} 
( c_{{\bm k}\uparrow}^\dagger, c_{-{\bm k}\downarrow} )
\left(\begin{array}{cc} \xi_{\bm k} & \Delta  \\
\Delta & -\xi_{\bm k} \end{array} \right)
\left( \begin{array}{c}
 c_{{\bm k}\uparrow}  \\
 c_{-{\bm k}\downarrow}^\dagger \end{array} \right), 
 \label{HSC} 
\end{equation}
where $c_{{\bm k}\sigma}$ ($c_{{\bm k}\sigma}^\dagger$) is the annihilation (creation) operator of the electrons in the superconductors, $\xi_{\bm k}$ is the energy of conduction electrons measured from the chemical potential.
The order parameter of the SC, $\Delta$, is determined by the gap equation
\begin{equation}
{\rm ln} \left(\frac{T}{T_{\rm c}}\right) \Delta = 2\pi T \sum_{\varepsilon_n >0 }
\left(\frac{\Delta}{\sqrt{\varepsilon_n^2+\Delta^2}} - \frac{\Delta}{\epsilon_n}\right),
\end{equation}
where $\varepsilon_n = (2n+1) \pi T$ is the Matsubara frequency, and $T_{\rm c}$ is the SC transition temperature~\cite{Inoue17}.

The second term $H_{\rm FI}$ describes a bulk FI, and is given by the Heisenberg model
\begin{eqnarray}
H_{\rm FI} &=& \sum_{\langle i,j \rangle}
J_{ij} {\bm S}_i \cdot {\bm S}_j - \hbar \gamma h_{\rm dc} \sum_{i} S^{z}_{i}
\nonumber \\
&-& \frac{\hbar \gamma h_{\rm ac}}{2}\sum_{i} (e^{-i\Omega t} S^{-}_i
+e^{i\Omega t} S^{+}_i ),
\label{HFI} 
\end{eqnarray}
where ${\bm S}_{i}$ is the localized spin at site $i$ in the FI, $J_{ij}$ is the exchange interaction, $h_{\rm dc}$ is a static magnetic field, $h_{\rm ac}$ and $\Omega$ are the amplitude and frequency of the applied microwave radiation, respectively, and $\gamma$ is the gyromagnetic ratio.
Using the Holstein-Primakoff transformation~\cite{Holstein40} and employing the spin-wave approximation ($S^z_j = S_0 - b_j^\dagger b_j$, $S^+_j \simeq (2S_0)^{1/2} b_j$), the Hamiltonian of the FI is rewritten as
\begin{eqnarray}
H_{\rm FI} &\simeq& {\rm const.}+ \sum_{\bm k} \hbar \omega_{\bm k} b_{\bm k}^\dagger b_{\bm k}
\nonumber \\
&-& 
\frac{\hbar \gamma h_{\rm ac}}{2} \sqrt{2S_0N_{\rm F}}  (e^{-i\Omega t} b_{\boldsymbol{k}=\boldsymbol{0}}^\dagger
+e^{i\Omega t} b_{\boldsymbol{k}=\boldsymbol{0}}),
\label{HFI2} 
\end{eqnarray}
where $\hbar \omega_{\bm k}$ is the magnon dispersion, $b_{\bm k}$ is the Fourier transform of $b_j$, $S_0$ is the magnitude of the localized spin, and $N_{\rm F}$ is the number of spins in the FI.
For simplicity, we assume a parabolic dispersion $\hbar \omega_{\bm k} = Dk^2 + E_0$, where 
$E_0 = \hbar \gamma h_{\rm dc}$ is the Zeeman energy.

The last term in the system Hamiltonian, $H_{\rm ex}$, describes the exchange coupling at the interface.
In this paper, we employ a simple model using the following tunneling Hamiltonian for spins:
\begin{equation}
H_{\rm ex} = \sum_{{\bm k},{\bm q}} \left[{\cal T}_{{\bm k},{\bm q}} S^{+}_{\bm k} s^{-}_{\bm q} 
+ {\rm h.c.} \right],
\label{Hex}
\end{equation}
where ${\cal T}_{{\bm k},{\bm q}}$ is the tunneling amplitude, $S^{+}_{\bm k} = (2S_0)^{1/2} b_{\bm k}$, and $s_{\bm q}^-$ is the operator defined as
\begin{equation}
s^{-}_{\bm q} := \sum_{{\bm k}} c^\dagger_{{\bm k} \downarrow} c_{{\bm k}+{\bm q} \uparrow}.
\end{equation}
In what follows, we study the spin transport by considering a second-order perturbative expansion with respect to $H_{\rm ex}$.

\section{Dynamic Spin Susceptibility}

\label{sec:SpinSusceptibility}

In this section, we summarize the results for the dynamic spin susceptibilities for the unperturbed system, i.e., the decoupled FI and SC, which are later used in the second-order perturbation calculation of the spin current and spin-current noise.

\subsection{Retarded component}

We define the retarded components of the spin susceptibility of the SC and the magnon propagator in the FI as
\begin{align}
& \chi^{R} ({\bm q},t) 
 := i(\hbar N_{\rm S})^{-1} \theta(t)\langle 
    [s^+_{{\bm q}}(t),s^-_{{\bm q}}(0)]
    \rangle \label{eq:retardedchi}, \\
& G^{R} ({\bm k},t) 
 := -i\hbar^{-1} \theta(t)\langle 
    [S^+_{{\bm k}}(t),S^-_{{\bm k}}(0)]
    \rangle \label{eq:retardedG}, 
\end{align}
where $N_{\rm S}$ is the number of unit cells in the SC.
Their Fourier transformations are defined as
\begin{align}
& \chi^R({\bm q},\omega) := \int_{-\infty}^\infty dt \, e^{i\omega t} \chi^R({\bm q},t), \\
& G^R({\bm k},\omega) := \int_{-\infty}^\infty dt \, e^{i\omega t} G^R({\bm k},t).
\end{align}

We first consider the magnon propagator of the FI.
By using the Holstein-Primakoff transformation~\cite{Holstein40} and employing the spin-wave approximation ($S^+_{\bm k} \simeq (2S_0)^{1/2} b_{\bm k}$), the magnon propagator of the FI is calculated in the absence of the external field ($h_{\rm ac}=0$) as
\begin{align}
G^{R}({\bm k},\omega) &= \frac{2S_0/\hbar}{\omega- \omega_{\bm k}+i\alpha \omega},
\label{eq:magnonpropagator}
\end{align}
where we have introduced the phenomenological dimensionless damping parameter $\alpha$, which originates from the Gilbert damping.

Next, we consider the dynamic spin susceptibility of the SC in the BCS theory.
We define the local spin susceptibility as
\begin{equation}
\chi^{R}_{\rm loc} (\omega) := \frac{1}{N_{\rm S}} \sum_{\bm q} \chi^{R}({\bm q},\omega).
\end{equation}
In the BCS theory, the local spin susceptibility is calculated as~\cite{Coleman15}
\begin{align}
{\rm Im} \chi^{R}_{\rm loc} (\omega) &= - \pi N(\epsilon_{\rm F})^2 \int dE \left[1 + \frac{\Delta^2}{E(E+\hbar \omega)}\right]\nonumber \\
& \times [f(E+\hbar \omega)-f(E)]D(E)D(E+\hbar\omega) ,
\label{eq:ImChiloc} \\
D(E) &= \frac{|E|}{\sqrt{E^2-\Delta^2}} \theta(E^2-\Delta^2),
\label{eq:SCDE}
\end{align}
where $N(\epsilon_{\rm F})$ is the density of states per spin and per unit cell, $f(E)=(e^{E/k_{\rm B}T}+1)^{-1}$ is the Fermi distribution function, $D(E)$ is the (normalized) density of states of quasi-particles, and
$\theta(x)$ is the Heaviside step function (for a detailed derivation, see Appendix~\ref{app:BCS}).
We note that the factor $[1+\Delta^2/E(E+\hbar \omega)]$ in Eq.~(\ref{eq:ImChiloc}) is the so-called coherence factor, which produces singular behavior near the transition temperature~\cite{Coleman15}.
For the normal metal ($\Delta=0$), the local spin susceptibility becomes
\begin{equation}
{\rm Im} \, \chi^{R}_{{\rm loc},\Delta=0} (\omega) = \pi N(\epsilon_{\rm F})^2 \hbar \omega.
\label{eq:ImChilocNormal}
\end{equation}

\subsection{Lesser component}

We define the lesser components of the spin susceptibilities for bulk SC and FI as 
\begin{eqnarray}
& & \chi^{<} ({\bm q},t) 
 :=  i(\hbar N_{\rm S})^{-1} \langle 
    s^-_{{\bm q}}(0) s^+_{{\bm q}}(t) 
    \rangle \label{eq:lesser-chi},\\
& & G^{<} ({\bm k},t) 
 :=  -i\hbar^{-1} \langle 
    S^-_{{\bm k}}(0) S^+_{{\bm k}}(t) 
    \rangle \label{eq:lesser-g}.
\end{eqnarray}
Their Fourier transformations are defined as
\begin{align}
& \chi^<({\bm q},\omega) = \int_{-\infty}^\infty dt  e^{i\omega t} \chi^<({\bm q},t), \\
& G^<({\bm k},\omega) = \int_{-\infty}^\infty dt  e^{i\omega t} G^<({\bm k},t).
\end{align}
The lesser components include the information on the distribution function; we define the distribution functions as
\begin{align}
& f^{\rm SC}({\bm q},\omega) :=\chi^{<}({\bm q},\omega)/(2i) {\rm Im} \, \chi^{R}({\bm q},\omega), \label{f1} \\
& f^{\rm FI}({\bm k},\omega) := G^{<}({\bm k},\omega)/(2i) {\rm Im} \, G^{R}({\bm k},\omega). \label{f2}
\end{align}

In the setup of the spin pumping (SP), the SC is in equilibrium with the temperature $T$, whereas magnons in FI are excited by the external microwave irradiation.
We split the Hamiltonian of the FI as $H_{\rm FI} = H_0 + V$, where
\begin{align}
& H_0 = \sum_{\bm k} \hbar \omega_{\bm k} b_{\bm k}^\dagger b_{\bm k}, \\
& V =  -h_{\rm ac}^{+}(t) b_0^\dagger - h_{\rm ac}^{-}(t) b_0,\\
& h_{\rm ac}^{\pm}(t) = \frac{\hbar \gamma h_{\rm ac}}{2} \sqrt{2S_0 N_F}e^{\mp i\Omega t}.
\end{align}
While the perturbation $V$ does not change the retarded component of the dynamic spin susceptibility of FI, it does modify the lesser component.
The second-order perturbation with respect to $V$ gives the correction:
\begin{align}
& \delta G^<({\bm k},\omega) = G_0^R({\bm k},\omega) \Sigma({\bm k},\omega) G_0^A({\bm k},\omega), \\
& \Sigma({\bm k},\omega) = \delta_{{\bm k},{\bm 0}} \int dt \, (-i\hbar^{-1}) \langle h_{\rm ac}^{-}(t) h_{\rm ac}^{+}(0) \rangle e^{i\omega t},
\end{align}
where $G_0^R({\bm k},\omega)$ is the unperturbed spin susceptibility of FI.
One can then straightforwardly obtain
\begin{align}
\delta f^{\rm FI}({\bm k},\omega) &=  \delta G^<({\bm k},\omega)/(2i){\rm Im}G_0^R ({\bm k},\omega) \nonumber \\
& = \frac{2\pi N_{\rm F} S_0 (\gamma h_{\rm ac}/2)^2}{\alpha \omega}
\delta_{{\bm k},{\bm 0}} \delta(\omega-\Omega).
\label{eq:deltafSP}
\end{align}

In the setup of the spin Seebeck effect (SSE), FI and SC are in equilibrium with temperatures, $T_{\rm FI}$ and $T_{\rm SC}$, respectively.
Using their Lehmann representation, we can prove the relations\cite{Bruus04,Stefanucci13}
\begin{eqnarray}
& & \chi^<({\bm q},\omega) = 2i \, {\rm Im}\, \chi^R({\bm q},\omega) \, n_{\rm B}(\omega,T_{\rm SC}), \label{eq:chidist} \\
& & G^<({\bm q},\omega) = 2i \, {\rm Im}\, G^R({\bm q},\omega) \, n_{\rm B}(\omega,T_{\rm FI}), \label{eq:Gdist}
\end{eqnarray}
where $n_{\rm B}(\omega,T)$ is the Bose distribution function defined as
\begin{equation}
n_{\rm B}(\omega,T) = \frac{1}{e^{\hbar \omega/k_{\rm B}T}-1}.
\end{equation}
This result leads to the distribution functions of the FI and the SC (defined in Eqs.~(\ref{f1}) and (\ref{f2})) as
\begin{eqnarray}
f^{\rm SC}({\bm q},\omega) & = & n_{\rm B}(\omega,T_{\rm SC}),
\label{eq:distSCeq}\\ 
f^{\rm FI}({\bm k},\omega) & = & n_{\rm B}(\omega,T_{\rm FI}). \label{eq:distFIeq}
\end{eqnarray}

\section{Spin Current}
\label{sec:SpinCurrent}

\subsection{Formulation}
\label{sec:SpinCurrentFormulation}

The spin current at the SC-FI interface is defined by $\langle \hat{I}_S \rangle$, where $\langle \cdots \rangle$ denotes the statistical average, and
$\hat{I}_S$ is the operator for the spin current flowing from the SC to the FI defined by
\begin{eqnarray}
& & \hat{I}_{\rm S} := -\hbar \, \partial_t \left( s^z_{\rm tot} \right) = i [s_{\rm tot}^z, H], \\
& & s^z_{\rm tot} := \frac{1}{2} \sum_{\bm k} (c_{{\bm k}\uparrow}^\dagger c_{{\bm k}\uparrow}-c_{{\bm k}\downarrow}^\dagger c_{{\bm k}\downarrow}).
\end{eqnarray}
By substituting the expression for the system Hamiltonian, we obtain
\begin{equation}
\hat{I}_S = -i \sum_{{\bm k},{\bm q}} ({\cal T}_{{\bm k},{\bm q}} S_{\bm k}^+ s_{\bm q}^- - {\rm h.c.}) .
\end{equation}

\begin{figure}[!tb]
\begin{center}
\includegraphics[width=6.5cm]{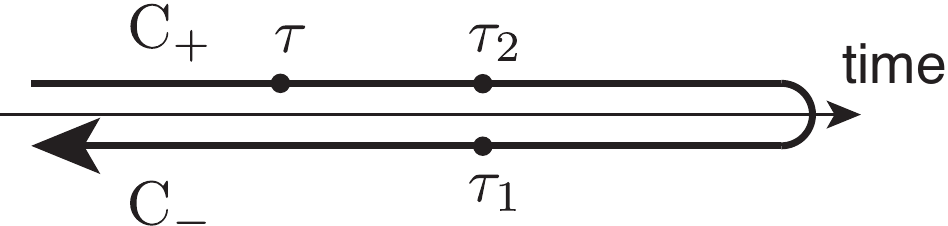}
\caption{The Keldysh contour C.}
\label{fig:Keldysh}
\end{center}
\end{figure}

We consider the second-order perturbation calculation by taking $H_{\rm FI}+H_{\rm SC}$ as an unperturbed Hamiltonian, and $H_{\rm ex}$ as a perturbation.
The average of the spin current operator is written as
\begin{align}
    \langle \hat{I}_S \rangle &= {\rm Re} \left[ -2i \sum_{{\bm k},{\bm q}} {\cal T}_{{\bm k},{\bm q}} \langle s_{\bm q}^- S_{\bm k}^+ \rangle \right] \nonumber \\
    & = \lim_{t_1, t_2\rightarrow 0} {\rm Re} \,  \left[ -2i \sum_{{\bm k},{\bm q}} {\cal T}_{{\bm k},{\bm q}} \langle s_{\bm q}^-(t_2) S_{\bm k}^+(t_1) \rangle \right],
\end{align}
where the average $\langle \cdots \rangle$ is taken for the full Hamiltonian.
By using the formal expression of perturbation expansion, the spin current can be rewritten as~\cite{Stefanucci13,Rammer86}
\begin{align}
    \langle \hat{I}_S \rangle 
    &= {\rm Re} \, \Biggl[ -2i \sum_{{\bm k},{\bm q}} {\cal T}_{{\bm k},{\bm q}} \langle T_{\rm K}  s_{\bm q}^-(\tau_2) S_{\bm k}^+(\tau_1) 
    \Biggr. \nonumber \\
    &\hspace{10mm} \times \Biggl. \exp \biggl(-\frac{i}{\hbar} \int_{\rm C} d\tau \, H_{\rm ex}(\tau) \biggr) \rangle_0 \Biggr],
    \label{eq:formal}
\end{align}
where the average $\langle \cdots \rangle_0$ is now taken for the unperturbed Hamiltonian, and $T_{\rm K}$ is the time-ordering operator on the time variable $\tau$ on the Keldysh contour C, which is composed of the forward path ${\rm C}_+$ running from $-\infty$ to $\infty$ and the backward path ${\rm C}_-$ from $\infty$ to $-\infty$ (see Fig.~\ref{fig:Keldysh}).
We have put the time variables, $\tau_1$ and $\tau_2$ on the contour ${\rm C}_-$ and ${\rm C}_+$, and have removed the limit operation for operator ordering.

Expanding the exponential operator in Eq.~(\ref{eq:formal}) and keeping the lowest-order term with respect to $H_{\rm ex}$, we obtain
\begin{align}
    \langle \hat{I}_S \rangle 
    &= -\frac{2}{\hbar} \int_{\rm C} d\tau \, {\rm Re}\, \Biggl[  \sum_{{\bm k},{\bm q}} |{\cal T}_{{\bm k},{\bm q}}|^2 \langle T_{\rm K} s_{\bm q}^+(\tau)s_{\bm q}^-(\tau_2) \rangle_0 \Biggr. \nonumber \\
    & \hspace{10mm} \times \Biggl.
    \langle T_{\rm K} S_{\bm k}^+(\tau_1) S_{\bm k}^-(\tau) \rangle_0 .
     \Biggr]
\end{align}
Using the real-time representation~\cite{Stefanucci13,Bruus04,Rammer86}, we can rewrite the spin current in terms of the dynamic spin susceptibilities of FI and SC as
\begin{align}
    \langle \hat{I}_S \rangle 
    &= - 2\hbar \, {\rm Re}\, \int_{-\infty}^{\infty} dt \, \sum_{{\bm k},{\bm q}} |{\cal T}_{{\bm k},{\bm q}}|^2 N_{\rm S} 
    \nonumber \\
    & \hspace{2.5mm} \times [ \chi^R({\bm q},t) G^<({\bm k},-t)
    + \chi^<({\bm q},t) G^A({\bm k},-t) ] ,
\end{align}
where $G^A({\bm k},t)$ is the advanced component. Using the definitions of the distribution functions and performing the Fourier transformation for the dynamic spin susceptibilities, we obtain
\begin{align}
\langle \hat{I}_S \rangle 
&= 4 \hbar \int \frac{d\omega}{2\pi} \sum_{{\bm k},{\bm q}} |{\cal T}_{{\bm k},{\bm q}}|^2 N_{\rm S} {\rm Im} \, \chi^R({\bm q},\omega)\nonumber \\
& \times (- {\rm Im}\, G^R({\bm k},\omega))
[ f^{\rm FI}({\bm k},\omega) - f^{\rm SC}({\bm q},\omega) ]
\end{align}
Setting ${\cal T}_{{\bm k},{\bm q}}={\cal T}$ for simplicity, we obtain
\begin{align}
 \langle \hat{I}_S \rangle 
&= \hbar A \int \frac{d(\hbar \omega)}{2\pi} \frac{1}{N_{\rm S}N_{\rm F}} \sum_{{\bm k},{\bm q}}{\rm Im} \, \chi^R({\bm q},\omega)\nonumber \\
&\times (- {\rm Im}\, G^R({\bm k},\omega))
[ f^{\rm FI}({\bm k},\omega) - f^{\rm SC}({\bm q},\omega) ], 
\end{align}
where $A = 4|{\cal T}|^2 N_{\rm S}^2 N_{\rm F}/\hbar$.

\subsection{Spin pumping}

We first consider the case of spin pumping driven by microwave irradiation keeping the same temperature for both SC and FI. 
From Eq.~(\ref{eq:deltafSP}), the difference of the distribution functions is given by
\begin{align}
& f^{\rm FI}({\bm k},\omega) - f^{\rm SC}({\bm q},\omega) 
\nonumber \\
& \hspace{10mm} = \frac{2\pi N_F S_0(\gamma h_{\rm ac}/2)^2}{\alpha \omega} 
\, \delta_{{\bm k},{\bm 0}}  \,
\, \delta (\omega -\Omega)\, .
\label{eq:distspinpumping}
\end{align}
The spin current generated by SP is then given by
\begin{align}
&I_S^{\rm SP} = \hbar \, A \, g(\Omega) \; {\rm Im} \chi^R_{\rm loc}(\Omega), \\
& g(\Omega) := \frac{(\gamma h_{\rm ac} S_0)^2/2}{(\Omega-\omega_0)^2+\alpha^2\Omega^2},
\end{align}
where the local spin susceptiblity $\chi^R_{\rm loc}(\omega)$ is given by Eqs.~(\ref{eq:ImChiloc}) and (\ref{eq:SCDE}), and $\omega_0 = \gamma h_{\rm dc}$ is the angular frequency of the spin precession.

For the normal metal case ($\Delta = 0$), we obtain for the spin current using Eq.~(\ref{eq:ImChilocNormal}):
\begin{equation}
I_{\rm S}^{\rm SP, N} = \pi \hbar A g(\Omega)N(\epsilon_{\rm F})^2 \hbar \Omega ,
\end{equation}
which is temperature independent for arbitrary values of $\Omega$. We will use this expression as a normalization factor to compare the results at finite $\Delta$ for various frequencies $\Omega$.

Before showing the results obtained in the superconducting case,
we point out that in the small frequency limit ($\Omega \to 0$), the expression for the spin current generated by SP is similar
to the one obtained when computing nuclear spin resonance (NMR) signal~\cite{Coleman15}.
It is known in the theory of the NMR measurement that the BCS singularity in the density of states leads to a coherence peak below the SC transition temperature~\cite{Hebel59,Masuda62}. As a consequence, one can expect a similar coherence peak in the temperature dependence of the spin current at low frequency. However the spin current contains more information than the NMR expression, since $\Omega$ can be controlled arbitrarily up to high frequencies of the order of the transition temperature $T_{\rm c}$. 

\begin{figure}[!tb]
\begin{center}
\includegraphics[width=8.cm]{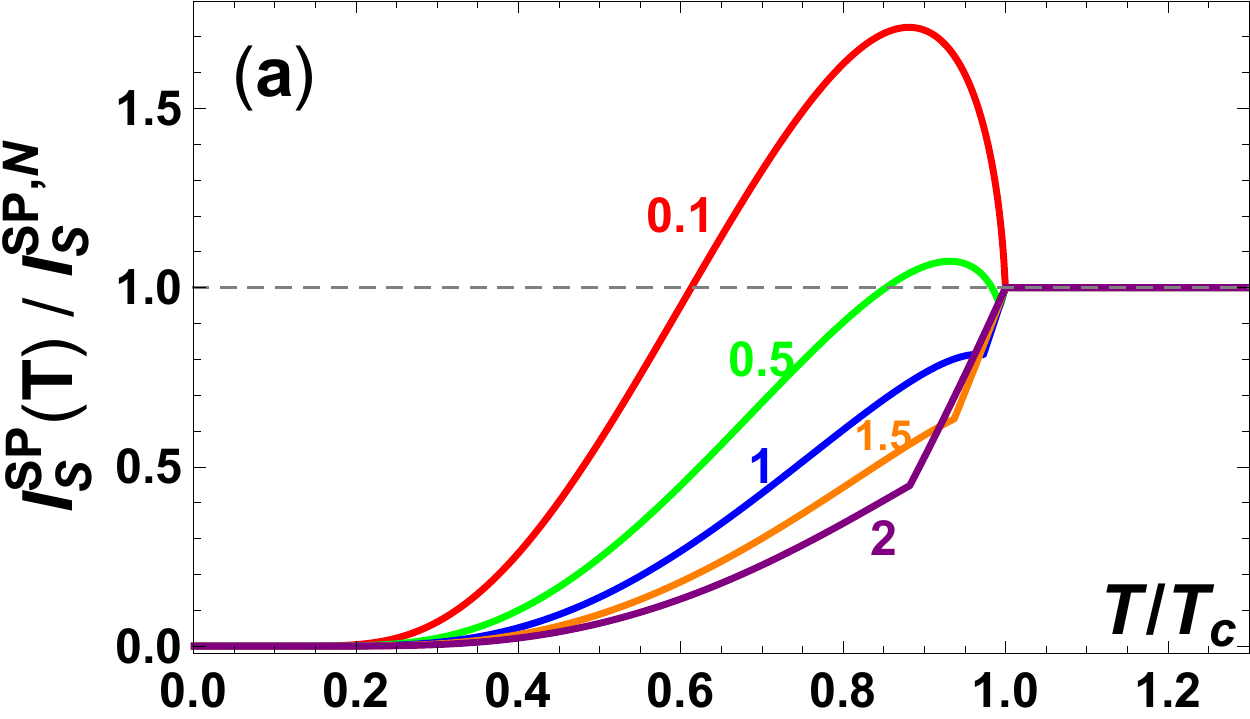}
\includegraphics[width=8.cm]{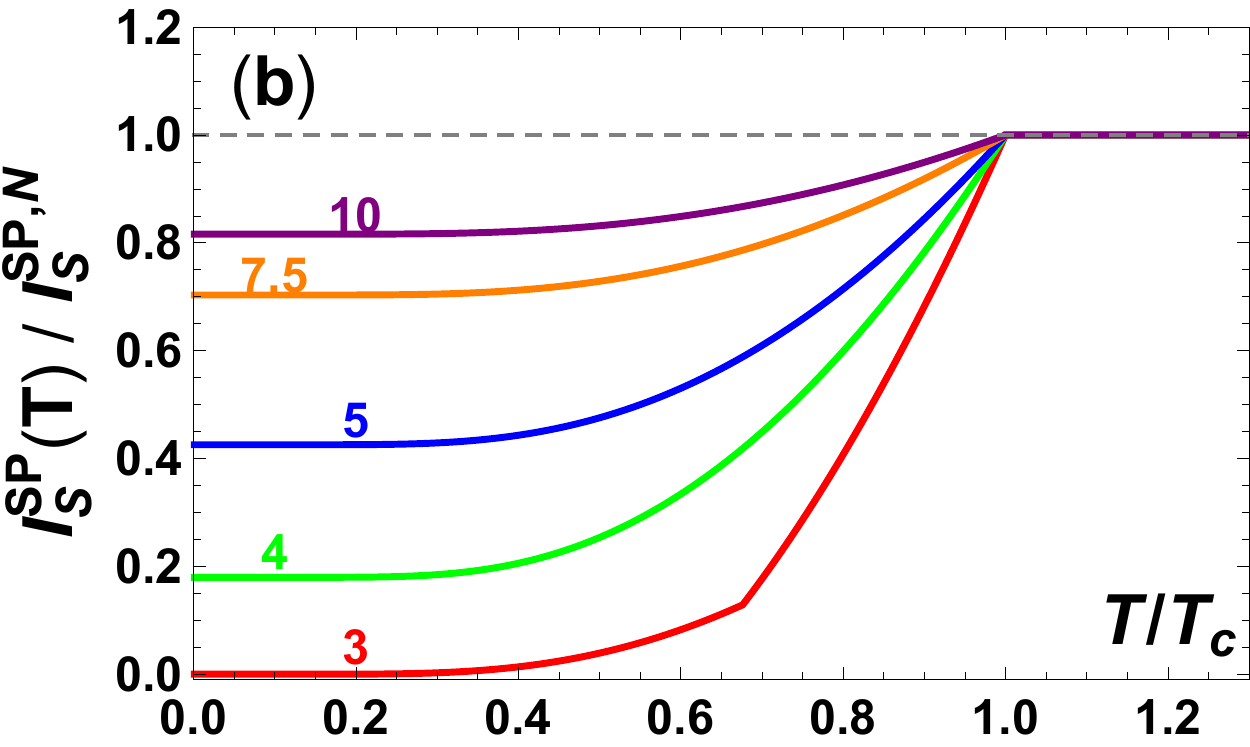}
\caption{Temperature dependence of the spin current induced by 
spin pumping $I^{SP}_S(T)$,
normalized by the current obtained in the normal case $I^{SP,N}_S(T)$, for different values of $\hbar \Omega/T_{\rm c}$, as indicated near each curve. Plot (a) shows  $\hbar \Omega/T_{\rm c}$ from 0.1 to 2. Plot (b) shows $\hbar \Omega/T_{\rm c}$ from 3.0 to 10.}
\label{fig:Isp}
\end{center}
\end{figure}

In Fig.~\ref{fig:Isp}, we show the temperature dependence of the spin current induced by spin pumping. Here the temperatures of both FI and SC are set to $T$, and the spin current is normalized by the value obtained for the normal metal case $I_{\rm S}^{\rm SP, N}$.
For small excitation frequency $\Omega$, the temperature dependence of $I_{\rm S}^{\rm SP}$ clearly shows a coherence peak below the SC transition temperature $T_{\rm c}$ as expected.
For $\hbar \Omega < 2\Delta(T=0) \simeq 3.54k_{\rm B}T_{\rm c}$, the spin current is strongly reduced at low temperatures ($k_{\rm B}T \ll 2\Delta(T)$), because spin-flip excitations in the SC are suppressed due to the energy gap $2\Delta$ in the one-electron excitation spectrum.
As $\Omega$ increases, the coherence peak becomes insignificant, while there appears a kink at the temperature satisfying $2\Delta(T) = \hbar \Omega$.
For $\hbar \Omega > 2\Delta(T=0)$, the spin current shows a plateau at low temperature corresponding to its zero temperature value, ultimately recovering the normal state value (dashed line) as $\hbar \Omega$ is increased further.

\subsection{Spin Seebeck effet}

We now turn to the alternative technique for generating a spin current, namely the spin Seebeck effect, which relies on the presence of a temperature gradient between the FI and the SC layers. 
Using Eqs.~(\ref{eq:distSCeq}) and (\ref{eq:distFIeq}),
the spin current induced by the spin Seebeck effect is given by
\begin{eqnarray}
& & I_{\rm S}^{\rm SSE} = \hbar A \int \!\! \frac{d(\hbar \omega)}{2\pi} \,
{\rm Im} \, \chi_{\rm loc}^R(\omega)  
(-{\rm Im} \, G^R_{\rm loc}(\omega)) \nonumber \\
& & \hspace{20mm} \times 
[n_{\rm B}(\omega,T_{\rm FI}) - n_{\rm B}(\omega, T_{\rm SC})],
\label{eq:Isexpression2} 
\end{eqnarray}
where $G^R_{\rm loc}(\omega):=N_{\rm F}^{-1} \sum_{{\bm k}} \, G^R({\bm k},\omega)$ is the local spin susceptibility in the FI.
For simplicity, we consider the spin Seebeck effect up to the linear term with respect to the temperature difference $\delta T = T_{\rm FI} - T_{\rm SC}$:
\begin{eqnarray}
& &\frac{I_{\rm S}^{\rm SSE}}{I_{{\rm S},0}^{\rm SSE}} = \int_{E_0}^{E_{\rm M}} dE\, D_{\rm M}(E) F(E)
\frac{(E/2k_{\rm B}T)^2}{\sinh^2(E/2k_{\rm B}T)}, \\
& & F(E) := {\rm Im} \chi^R_{\rm loc}(E/\hbar)/{\rm Im}\chi^R_{{\rm loc},\Delta=0}(E/\hbar) \nonumber \\
& & \hspace{2.5mm} =
\int_{-\infty}^{\infty} dE' \left[ 1+\frac{\Delta^2}{E'(E'+E)} \right] \nonumber \\
& & \hspace{5mm} \times \left[\frac{f(E')-f(E'+E)}{E}\right]
D(E') D(E'+E),
\end{eqnarray}
where $T = T_{\rm SC} \simeq T_{\rm FI}$ and $I_{{\rm S},0}^{\rm SSE}= \hbar A S_0 k_{\rm B} \delta T N(\epsilon_{\rm F})^2$. 
The density of states per site for magnon is given by
\begin{eqnarray}
D_{\rm M}(E) &:=& \frac{1}{N_{\rm F}} \sum_{\bm k} \delta(E-\hbar \omega_{\bm k}) \nonumber \\
&=& -(2\pi S_0)^{-1} {\rm Im} \, G_{\rm loc}^R(E/\hbar),
\end{eqnarray}
taking the limit $\alpha \rightarrow 0$, and $E_{\rm M}$ ($\gg E_0$) is the high-energy cut-off of the magnon dispersion relation, which is of the order of the exchange interaction in the FI.
Under a uniform magnetic field, the local spin susceptibility is evaluated for the parabolic magnon dispersion as $D_{\rm M}(E) = (3/2)(E-E_0)^{1/2}E_{\rm M}^{-3/2}$.
For normal metals ($\Delta =0$), the spin current at low temperatures ($k_{\rm B}T \ll E_{\rm M}$) is given by
$I_{\rm S}^{\rm SSE}/I_{{\rm S},0}^{\rm SSE}= \eta (k_{\rm B}T/E_{\rm M})^{3/2}$, where $\eta \simeq 6.69$ is a numerical factor.

\begin{figure}[tb]
\begin{center}
\includegraphics[width=8cm]{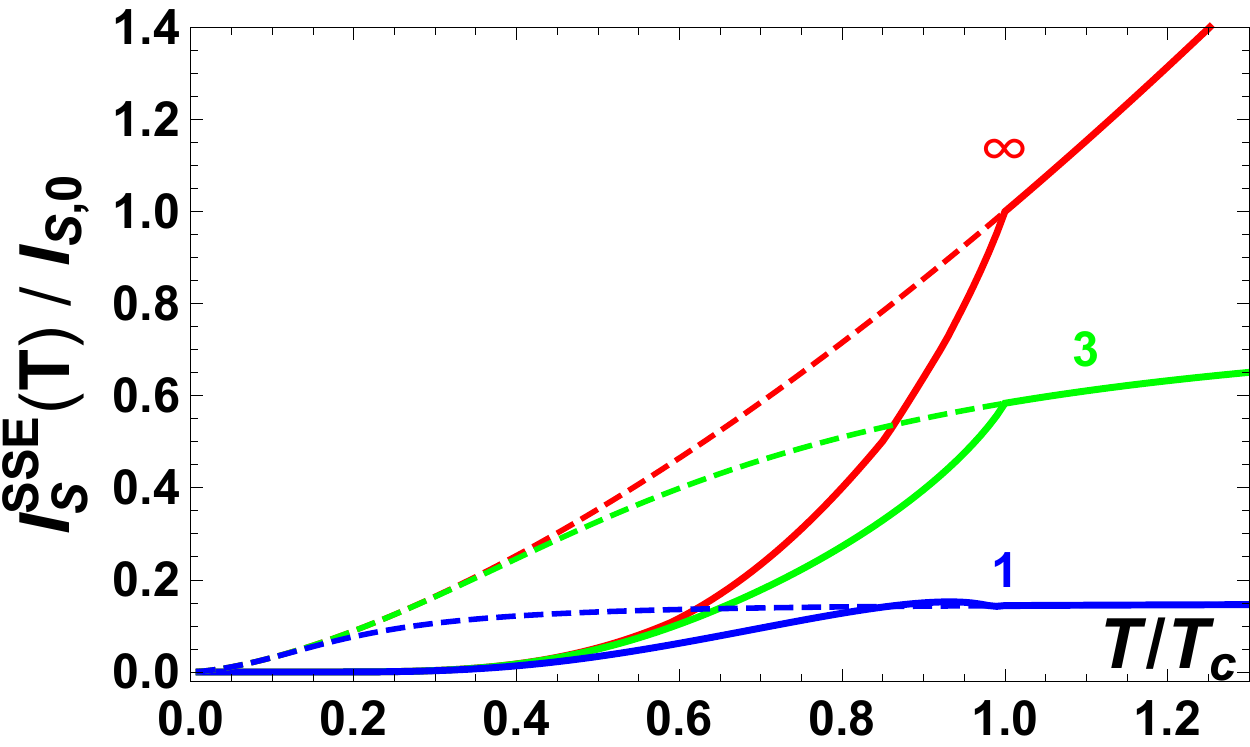}
\caption{Temperature dependence of the spin current induced by the spin Seebeck effect $I_{\rm S}^{\rm SSE}$, normalized by $I_{{\rm S},0}=I_{\rm S,0}^{\rm SSE} \eta (k_{\rm B}T_{\rm c}/E_{\rm M})^{3/2}$, for SCs (solid lines) and normal metals (dashed lines) with $E_{\rm M}/k_{\rm B}T_{\rm c} = \infty$, 3, and 1 (as indicated near each curve), where $E_{\rm M}$ is the high-energy cut-off of the magnon density of states. As we consider $E_0 \ll k_{\rm B} T_{\rm c}$, the Zeeman energy $E_0$ is set to zero for simplicity.}
\label{fig:Isse}
\end{center}
\end{figure}

In Fig.~\ref{fig:Isse}, we show the temperature dependence of $I_{\rm S}^{\rm SSE}$.
The solid and dashed lines show $I_{\rm S}^{\rm SSE}$ for the SC and the normal metal ($\Delta = 0$), respectively.
For simplicity, the Zeeman energy is set to zero by assuming that it is much smaller than $k_{\rm B}T$.
When $E_{\rm M}$ is much larger than $k_{\rm B}T_{\rm c}$, the spin current monotonically decreases as the temperature is lowered.
Below the transition temperature $T_{\rm c}$, the spin current at the FI-SC interface is suppressed due to the opening of the energy gap in the SC.
When $E_{\rm M}$ is comparable to $k_{\rm B}T_{\rm c}$, the spin current shows a small maximum below $T_{\rm c}$, and saturates above $T_{\rm c}$.

\section{Spin-Current Noise}
\label{sec:SpinCurrentNoise}

The noise of the pure spin current has been studied for an interface between a FI and a nonmagnetic metal~\cite{Matsuo18,Kamra16a,Kamra16b} as well as for several hybrid nanostructures~\cite{Aftergood17,Joshi18,Nakata18,Aftergood18}. 
It includes useful information on spin transport, as suggested from studies of the (electronic) current noise~\cite{ShotNoiseReview}.
In this section, we calculate the spin-current noise for the FI-SC interface.

\subsection{Formulation}

The noise power of the pure spin current is defined as~\cite{Matsuo18}
\begin{equation}
{\cal S} := \lim_{{\cal T}\rightarrow \infty} \frac{1}{{\cal T}}
\int_0^{{\cal T}} \! \! dt_1 \int_0^{{\cal T}} \! \! dt_2 \,
\frac{1}{2} \langle \{ \hat{I}_S(t_1), \hat{I}_S(t_2)\} \rangle,
\end{equation}
where $\hat{I}_S(t) := e^{iHt} I_S e^{-iHt}$, and $\{A,B\}=AB+BA$.
The spin-current noise is calculated within the second-order perturbation calculation with respect to $H_{\rm ex}$ as
\begin{align}
{\cal S} &= \hbar^2 \int_{-\infty}^{\infty}\frac{d\omega}{2\pi} \sum_{{\bm k},{\bm  q}}
 |{\cal T}_{{\bm k},{\bm q}}|^2 N_{\rm S} [\chi^<({\bm q},\omega) G^>({\bm k},\omega) \nonumber \\
 & \hspace{10mm} + \chi^>({\bm q},\omega) G^<({\bm k},\omega) ].
\end{align}
Using ${\cal T}_{{\bm k},{\bm q}}={\cal T}$, Eqs.~(\ref{f1})-(\ref{f2}), and the relations
\begin{align}
& \chi^{>}({\bm q},\omega)/(2i) {\rm Im} \, \chi^{R}({\bm q},\omega) = 1+f^{\rm SC}({\bm q},\omega) ,\\
& G^{>}({\bm k},\omega)/(2i) {\rm Im} \, G^{R}({\bm k},\omega) = 1+f^{\rm FI}({\bm k},\omega) ,
\end{align}
the spin-current noise is calculated as
\begin{align}
&{\cal S} = \hbar^2 A \int \frac{d(\hbar \omega)}{2\pi} \frac{1}{N_{\rm F}N_{\rm S}}\sum_{{\bm k},{\bm  q}}
 (-{\rm Im} G^R({\bm k},\omega)) {\rm Im} \chi^R({\bm q},\omega)\nonumber \\
& \hspace{10mm} \times [ f^{\rm SC}({\bm q},\omega)(1+ f^{\rm FI}({\bm k},\omega) ) \nonumber \\
& \hspace{15mm} + (1+f^{\rm SC}({\bm q},\omega))f^{\rm FI}({\bm k},\omega)].
\end{align}

In the absence of both the external microwave excitation and the temperature gradient, the noise power is determined by the equilibrium noise:
\begin{equation}
{\cal S}^{\rm eq} = 2 \hbar^2 A \int_{-\infty}^{\infty}\frac{d(\hbar \omega)}{2\pi} \frac{
  {\rm Im} \chi_{\rm loc}^R(\omega)(-{\rm Im} G_{\rm loc}^R(\omega))}{4\sinh^2(\hbar \omega/2k_{\rm B}T)}.
\end{equation}
Under the microwave radiation, the noise power is calculated from Eq.~(\ref{eq:distspinpumping}) as
\begin{eqnarray}
& & {\cal S} = {\cal S}^{\rm eq} + {\cal S}^{\rm SP}, \\
& & {\cal S}^{\rm SP} = \hbar \coth \left(\frac{\hbar \Omega}{2k_{\rm B}T}\right) I_{\rm S}^{\rm SP},
\end{eqnarray}
where ${\cal S}^{\rm SP}$ is the non-equilibrium noise induced by spin pumping. 
While the non-equilibrium noise can similarly be induced by SSE, we do not discuss it here as it requires to consider a large temperature gradient.

\subsection{Estimate}

\begin{figure}[tb]
\begin{center}
\includegraphics[width=8cm]{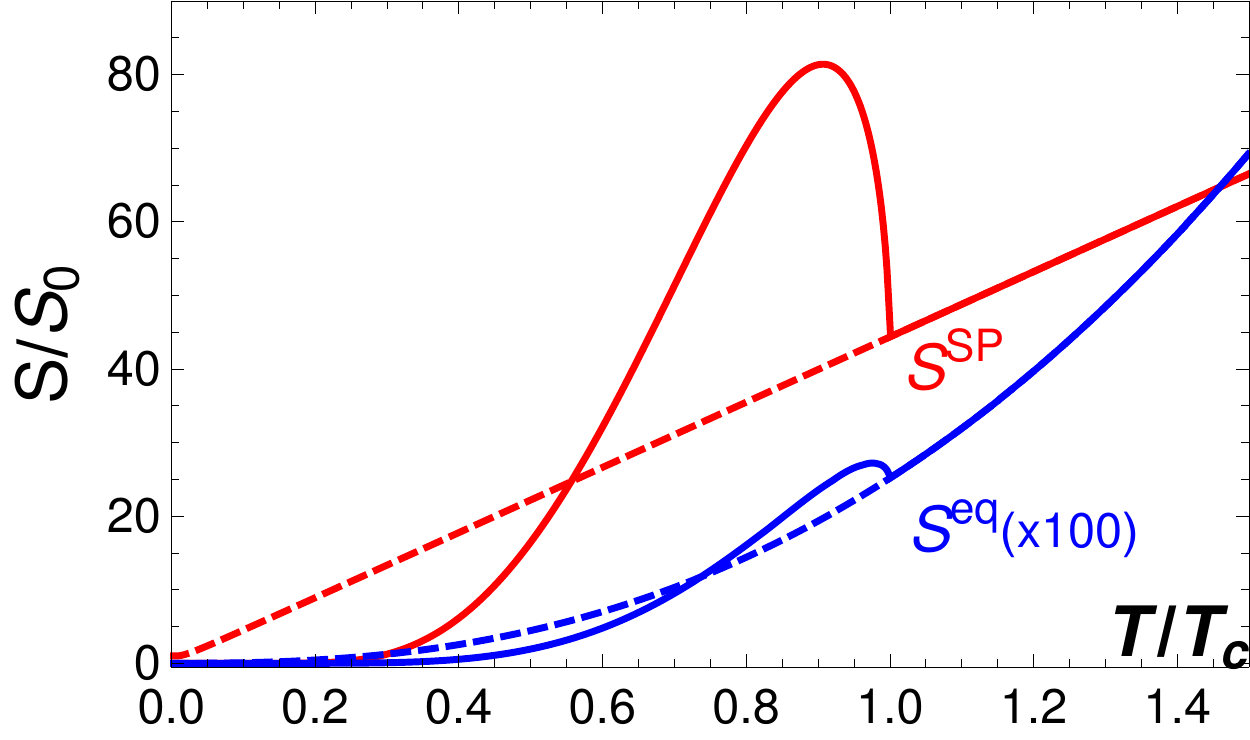}
\caption{Temperature dependence of the equilibrium noise ${\cal S}^{\rm eq}$ and the non-equilibrium noise in the spin pumping case ${\cal S}^{\rm SP}$ for SCs (solid lines) and normal metals (dashed lines).
The noise power is normalized by the non-equilibrium noise in the spin pumping case ${\cal S}_{\rm 0}$ for the normal metals at $T=0$.
For better visualization, data of the equilibrium noise has been multiplied by 100.
}
\label{fig:noise}
\end{center}
\end{figure}

As in the metal-FI bilayer system~\cite{Matsuo18,Kamra16a,Kamra16b}, the noise power of the pure spin current includes useful information also in the case of the SC-FI interface.
At low temperatures ($k_{\rm B}T\ll \hbar \Omega$), the ratio ${\cal S}^{\rm SP}/I_{\rm S}^{\rm SP}$ approaches $\hbar$, indicating that each magnon excitation carries the angular momentum $\hbar$.
At high temperatures ($k_{\rm B}T\gg \hbar \Omega$), this ratio becomes proportional to $k_{\rm B} T$ due to the nature of the Bose statistics.
To illustrate their temperature dependence, we estimate and compare the noise powers, ${\cal S}^{\rm eq}$ and ${\cal S}^{\rm SP}$, in realistic experiments.
We use the parameters of the spin pumping experiment for YIG~\cite{Kajiwara10}; $\alpha = 6.7\times 10^{-5}$, $S_0 = 16$, $h_{\rm ac} = 0.11\, {\rm Oe}$, $\gamma = 1.76\times 10^7 \, {\rm Oe}^{-1}{\rm s}^{-1}$ and $\Omega/2\pi = 9.4 \, {\rm GHz}$.
We consider NbN for the SC ($T_{\rm c} \simeq 10\, {\rm K}$), and set $D=532\, {\rm meV}$\AA${}^2$ following Ref.~\onlinecite{Princep17}.
Fig.~\ref{fig:noise} shows the results for the noise power, normalized by ${\cal S}_{\rm 0} = {\cal S}^{\rm SP}(T=0)$ for normal metals.
For this parameter set, the non-equilibrium noise associated with spin pumping is much larger than the equilibrium noise.
For both ${\cal S}^{\rm eq}$ and ${\cal S}^{\rm SP}$, the temperature dependence is peaked below the superconducting transition temperature.

\section{Experimental Setup for Detection}
\label{sec:experiment}

In the previous sections, we have evaluated the spin current and its noise at the FI-SC interface.
For their experimental detection, we need to consider a setup for converting the spin imbalance induced by the spin current into a charge signal.
There are several ways to perform such a spin-charge conversion.
Here, we explain one possible way using the inverse spin Hall effect (ISHE).
It was theoretically predicted that such spin current flowing in SC can be detected by the ISHE~\cite{Takahashi02,Takahashi08}.
Indeed, a giant signal of ISHE has recently been observed by spin injection from ferromagnetic metals into an $s$-wave superconductor NbN using the technique of the lateral spin valve~\cite{Wakamura15}.

\begin{figure}[tb]
\begin{center}
\includegraphics[width=7.5cm]{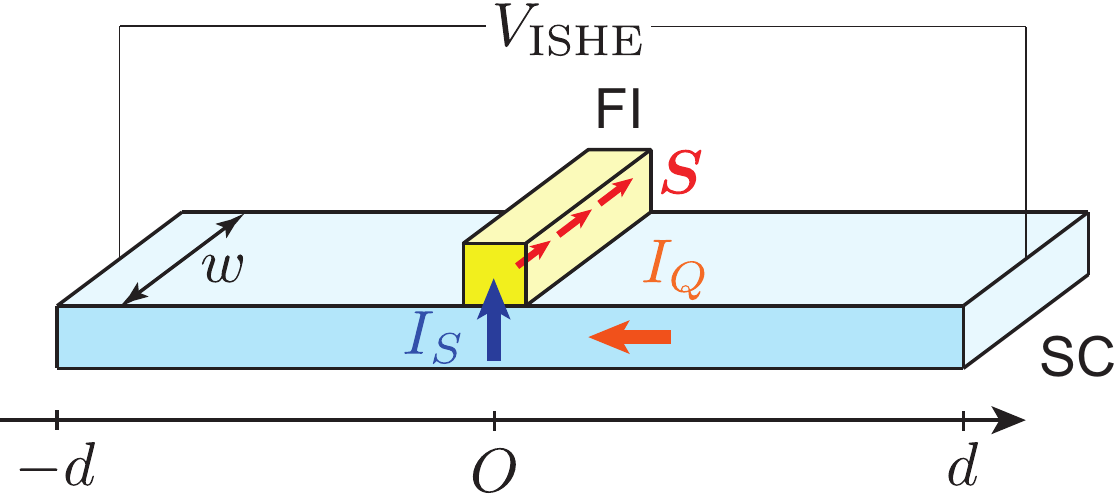}
\caption{A setup for detection of the spin current using the inverse spin Hall effect.}
\label{fig:ISHE}
\end{center}
\end{figure}

Let us consider spin injection into a SC wire with a width $w$ and a length $2d$ from a FI at $x=0$ (see Fig.~\ref{fig:ISHE}).
By spin-orbit interaction in the SC, the spin current $I_S$ is converted into a quasi-particle current $I_Q$ in the direction perpendicular to both $I_S$ and the ordered spin in the FI, ${\bm S}$.
This quasi-particle current induces a charge imbalance in the SC, and produces a voltage between the two edges located at $x=\pm d$.
Amplitude of the ISHE voltage depends on the spin relaxation in the SC as well as the spin-Hall angle, so that the coefficient between the spin current at the interface and the ISHE voltage is in general temperature-dependent.
Here, we introduce a simple formula employed in Ref.~\onlinecite{Wakamura15}:
\begin{align}
& V_{\rm ISHE} = \frac{|e|}{\hbar} I_S \frac{x}{w} \left(a \frac{\rho_{xx}}{2f_0(\Delta)} + b \biggl(\frac{\rho_{xx}}{2f_0(\Delta)} \biggr)^2 \right) e^{-d/\lambda_Q}, 
\label{eq:ISHE} \\
& f_0(\Delta) = \frac{1}{e^{\Delta/k_{\rm B}T}+1}. \label{eq:ISHE2}
\end{align}
This expression for the ISHE voltage has been derived assuming an extrinsic spin Hall effect due to spin-orbit scattering in the SC.
Here, $\lambda_Q$ is a charge relaxation length, $a$ and $b$ are coefficients determined by strength of skew scattering and side jump, respectively, and $\rho_{xx}$ is the resistivity of the SC.
Correction due to non-uniform current distribution is represented by a shunting length $x$, which is determined by $w$, $\lambda_Q$, and the shape of the junction~\cite{Wakamura15}.
Combining Eqs.~(\ref{eq:ISHE})-(\ref{eq:ISHE2}) with careful determination of the parameters, we can obtain $I_S$ from the measurement of $V_{\rm ISHE}$.
In principle, the spin-current noise can also be measured within the same kind of setup via the fluctuations of $V_{\rm ISHE}$~\cite{Matsuo18}.

\section{Summary}
\label{sec:summary}

In summary, we discussed the spin current and the spin-current noise for the bilayer system composed by a superconductor and a ferromagnetic insulator.
The spin current induced by spin pumping has a maximum below the transition temperature when the pumping frequency $\Omega$ is much smaller than $k_{\rm B}T_c/\hbar$.
As the ratio $\hbar \Omega/k_{\rm B}T_c$ increases, the peak disappears, and the spin current at low temperatures is enhanced.
We also discussed the spin current induced by spin Seebeck effect and the noise power of the pure spin current.
Our study provides a fundamental basis for the application of spintronics using superconductors.
Extension to spin injection from antiferromagnetic insulators is left for a future problem~\cite{Seki15,Wu16,Baltz18,Lado18}.

The authors are grateful to S. Takei, Y. Niimi, Y.-C. Otani, K. Kobayashi, and T. Arakawa for useful discussions and comments. 
This work is financially supported by ERATO-JST (JPMJER1402), and KAKENHI (Nos. 26103006, JP26220711, JP16H04023, and JP17H02927) from MEXT and JSPS, Japan. 
This work has been supported by the Excellence Initiative of Aix-Marseille University - A∗MIDEX, a French ``investissements d'avenir'' program.

\appendix

\section{Effect of Impurity Scattering}
\label{app:ImpurityEffect}

Here, we explain that the diffusive behavior of conduction electrons, which is taken into account in Ref.~\onlinecite{Inoue17}, can be neglected in the calculation of ${\rm Im} \chi^{R}_{\rm loc} (\omega)$ following Ref.~\onlinecite{Shastry94}.
We neglect Coulomb interaction effect discussed in Ref.~\onlinecite{Shastry94} for simplicity.
For a qualitative discussion, it is convenient to start with the interpolation formula (Eq.~(6) in Ref.~\onlinecite{Shastry94})
\begin{equation}
\chi_D^R({\bm q},\omega) \simeq
\chi_0({\bm q},\omega) \frac{{\cal D}q^2}{{\cal D}q^2-i\omega},
\end{equation}
where $q=|{\bm q}|$, $\chi_0({\bm q},\omega)$ is the spin susceptibility per volume of the electron gas,
${\cal D}=v_F l/3$ is the diffusion constant, $l=v_F \tau$ is the mean free path, $v_{\rm F}$ is the Fermi velocity, and $\tau$ is a relaxation time.
The leading behavior for small $\omega$ is (see Eq.~(7)  in Ref.~\onlinecite{Shastry94}) 
\begin{equation}
\frac{{\rm Im}\, \chi({\bm q},\omega)}{\hbar \omega} = \frac{N(\epsilon_{\rm F})}{\hbar} \left(
\frac{\pi}{2 v_F q} + \frac{1}{{\cal D}q^2} \right),
\quad (0<q<2k_F),
\end{equation}
where $k_{\rm F}$ is the Fermi wavenumber.
Then, the local spin susceptibility is calculated as
\begin{align}
\frac{{\rm Im} \, \chi_{\rm loc}(\omega)}{\hbar \omega} 
&= \int \frac{d^3{\bm q}}{(2\pi)^3} \frac{{\rm Im}\, \chi({\bm q},\omega)}{\hbar \omega} \nonumber \\
&= 2 \pi N(\epsilon_F)^2 \left( \frac12 + \frac{3}{\pi k_F l} \right)
\end{align}
Since $k_F l \gg 1$ for usual metals, the second term due to diffusive Green's function is usually a correction.
Therefore, the leading contribution is obtained only by considering a clean system without impurities.
For superconductors, a similar discussion leads to the same conclusion.

\section{Spin Susceptibility of the SC}
\label{app:BCS}

The dynamic spin susceptibility of the SC is calculated in the standard BCS theory as~\cite{Coleman15}
\begin{align}
\chi^{R}({\bm q},\omega)  &= \frac{1}{N_{\rm S}} \sum_{{\bm k}} \sum_{\lambda=\pm 1} \sum_{\lambda'=\pm 1}
\left[ \frac{1}{4}+ \frac{\xi \xi'+\Delta^2}{4E_{\lambda} E_{\lambda'}'} \right] \nonumber \\
& \times \frac{f(E_{\lambda'}')-f(E_{\lambda})}{\hbar \omega + i\delta + E_{\lambda} - E_{\lambda'}'},
\end{align}
where $\xi = \xi_{{\bm k}}$, $\xi' = \xi_{{\bm k}+{\bm q}}$, $E_\lambda = \lambda \sqrt{\Delta^2+\xi^2}$, $E_{\lambda'}' = \lambda' \sqrt{\Delta^2 + \xi'^2}$, and $f(E)=(\exp(E/k_{\rm B}T)+1)^{-1}$ is the Fermi distribution function.
For the normal state $(\Delta = 0)$, the spin susceptibility becomes
\begin{equation}
\chi^{R}({\bm q},\omega) = \frac{1}{N_{\rm S}} \sum_{{\bm k}}
\frac{f(\xi_{{\bm k}+{\bm q}})-f(\xi_{\bm k})}{\hbar \omega + i\delta + \xi_{\bm k} - \xi_{{\bm k}+{\bm q}}}.
\end{equation}
The imaginary part of the local spin susceptibility is obtained for the SC as
\begin{align}
{\rm Im} \chi^{R}_{\rm loc} (\omega) &= 
- \frac{\pi}{N_{\rm S}^2} \sum_{{\bm k},{\bm k}'}
\sum_{\lambda,\lambda'} 
\left[ \frac{1}{4}+ \frac{\xi \xi'}{4E_{\lambda} E_{\lambda'}'} \right]
\nonumber \\
& \times [f(E_{\lambda'}')-f(E_{\lambda})] \delta(\hbar \omega + E_{\lambda} - E_{\lambda'}'),
\end{align}
where $\xi'=\xi_{{\bm k}'}$ and $E_{\lambda'}'=\lambda' E_{{\bm k}'}$.
For $\hbar \omega \ll \epsilon_{\rm F}$ ($\epsilon_{\rm F}$: the Fermi energy), we can replace the wavenumber summation according to 
\begin{equation}
\frac{1}{N_{\rm S}} \sum_{\bm k} (\cdots) \rightarrow N(\epsilon_{\rm F}) \int_{-\infty}^{\infty} d\xi (\cdots), 
\end{equation}
where $N(\epsilon_{\rm F})$ is the density of states per spin and per unit cell.
Changing the integral variable from $\xi$ to $E=\sqrt{\Delta^2+\xi^2}$, we finally obtain Eqs.~(\ref{eq:ImChiloc}) and (\ref{eq:SCDE}).


\end{document}